\documentclass[a4paper,english,twocolumn]{revtex4}
\usepackage{times}
\usepackage[OT1]{fontenc}
\usepackage[latin1]{inputenc}
\usepackage{amsmath}
\usepackage{graphicx}
\usepackage{color}
\usepackage{amssymb}

\makeatletter
\usepackage{babel}
\makeatother
\begin{document}
 
\title{Classical Region of a Trapped Bose Gas}

\author{P. Blair Blakie$^{1}$ and Matthew J. Davis$^{2}$}

\affiliation{$^{1}$Jack Dodd Centre for Photonics and Ultra-cold Atoms, University of Otago, Dunedin,
New Zealand\\
\noindent$^{2}$ARC Centre of Excellence for Quantum-Atom Optics,
School of Physical Sciences, University of Queensland, Brisbane, QLD
4072, Australia}

\date{\today}

\pacs{03.75.-b, 03.75.Hh}

\begin{abstract}

The classical region of a Bose gas consists of all single-particle
modes that have a high average occupation and are well-described by
a classical field.
 Highly-occupied modes only occur in massive Bose gases at ultra-cold
temperatures, in contrast to the photon case where there are
highly-occupied modes at all temperatures. For the Bose gas the number of these
modes is dependent on the temperature, the total number of particles and their
interaction strength. In this paper we characterize the classical
region of a harmonically trapped Bose gas over a wide parameter regime.  We use
a Hartree-Fock approach to account for the effects of interactions, which we
observe to significantly change the classical region  as compared to the
idealized case.  We compare our results
to full classical field calculations and show that the Hartree-Fock approach
provides a qualitatively accurate description of classical region for the
interacting gas.

\end{abstract}
\maketitle

\section{Introduction}

There has been much recent work on using the classical field approximation
to model Bose-Einstein condensates at zero and finite temperatures
\cite{Davis2002a,DavisTemp,Davis2005a,Davis2006a,Simula2006a,Gardiner2002a,Gardiner2003a,Goral2001a,
Davis2001b,Davis2001a,Lobo2004a,Marshall1999a,Norrie2004a,
Polkovnikov2004a,Sinatra2001a,Sinatra2002a,Steel1998a,Bradley,Blakie2004}.
At zero temperature a single mode of the system (the condensate) is
macroscopically occupied and its evolution is well-described by the
Gross-Pitaevskii equation. The classical field approximation is also
suited to finite temperature regimes where many modes of the system
are highly occupied (also see \cite{Svistunov1991}).  The major advantage of classical field approaches
over other methods is that interactions between the modes can be treated
non-perturbatively, making the formalism suitable for non-equilibrium
studies.

The purpose of this paper is to fully characterize the size and nature of the classical region for typical experimental parameters. These properties of the classical region are not \emph{a priori} obvious, especially for the case of an interacting gas. Better characterization of the classical region will be useful for: (a) informing better choice of parameters for classical field simulations of finite temperature gases; and (b) demonstrating the regimes of applicability of this theory.

\subsection{Historic Motivation: Blackbody Radiation }

\begin{figure}
\includegraphics[%
  width=8cm,
  keepaspectratio]{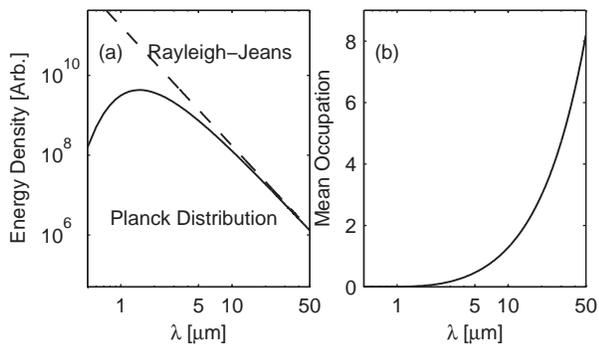}

\caption{\label{Fig:Blackbody} (a) Quantum and classical theories for
the energy spectrum of a blackbody at $T=2500$ K. (b) The mean number 
of photons in each mode at $T=2500$ K calculated using the Planck distribution.}
\end{figure}

We begin by reviewing a topic of modern physics that was responsible
for the quantum revolution, yet also serves as motivation for the
application of the classical field approximation to quantum fields:
the spectrum of radiation from a blackbody. Approximately a century
ago the Rayleigh-Jeans theory of the light spectrum emitted from a
blackbody was developed upon a classical field theory (i.e.\ Maxwell's
electromagnetic theory) in conjunction with statistical methods. The
famous failure of this theory eventually 
led to the discovery that light is constituted
of photons, and initiated the quantum revolution in physics. However,
an observation of prime importance is that the Rayleigh-Jeans theory
is actually extremely successful at predicting the behaviour of the
blackbody spectrum for the long wavelength modes. We illustrate
this point in Fig.~\ref{Fig:Blackbody}(a), where the quantum
and classical predictions for the spectrum are compared for a blackbody
at a temperature of $T=2500$ K. The most distinctive feature of this
graph is the rather alarming disagreement at small wavelengths, the
so-called ultra-violet catastrophe. However  at larger wavelengths, greater than $5\mu$m
say, the agreement between the quantum and classical predictions becomes
increasingly good. For these long wavelength modes each photon carries
only a small amount of energy compared to $k_BT$, so that on average
these modes contain  a large number of photons. This high
occupancy is what we will refer to as the classical limit for a field.
 For these
modes the discreteness of the energy each photon carries is masked
by the large number of other photons in the same mode and the classical
theory provides a good description.
We note that this is quite different
from the \emph{high temperature} classical limit for particles, when Bose/Fermi statistics can be
approximated by Maxwell-Boltzmann statistics.
The ultra-violet catastrophe occurs
because the classical theory fails to describe the short wavelength
modes correctly. These modes are sparingly occupied by photons and
require a proper quantum treatment. In Fig.~\ref{Fig:Blackbody}(b)
the mean number of photons per mode is shown, and comparison with
Fig.~\ref{Fig:Blackbody}(a) clearly reveals that the quantum
and classical theories agree well where the mean number of photons
is large. In general any electromagnetic mode can be made classical
by making the temperature sufficiently large  as photon number is not
conserved.

\subsection{Classical Region of a Bose Gas}

Atoms are necessarily conserved and require a chemical potential ($\mu$) in
their statistical description. 
By expanding the Planck and Bose-Einstein distributions  for high occupation $n(\epsilon) \gg 1$  we find the following requirements on the single particle energies 
\begin{eqnarray}
\epsilon &\ll&k_BT\qquad \rm{Planck},\\
\epsilon-\mu &\ll&k_BT\qquad \rm{Bose-Einstein}. 
\end{eqnarray}
We note that in these limits the classical equipartition theorem holds \footnote{For highly occupied modes of a Bose gas we obtain [to first order in the small parameter $(\epsilon-\mu)/k_BT$] the equipartition result for the mean occupation $n(\epsilon)=k_BT/(\epsilon-\mu)$.
}.
For the Bose gas at high temperatures (the Maxwell-Boltzmann limit) the chemical
potential is large and negative
\begin{eqnarray}
\mu \approx -k_B T \ln\left[\frac{V}{N}\left(\frac{2\pi m k_BT}{h^2}\right)^{3/2}\right],
\end{eqnarray}
where $V$ is the system volume and $N$ is the number of atoms. In contrast to the photon case, this behaviour of $\mu$ prevents the mean occupation of any individual mode from becoming large as $T$ increases.  

Indeed, for a gas of atoms at temperatures above
the microkelvin regime, the average occupation of the system modes
is much less than unity and the particle-like behaviour of the system
dominates over the wave-like behaviour. However, at temperatures approaching
the condensation temperature ($T_{c}$), the chemical potential approaches
the ground state energy, and many modes of the system may become highly
occupied. The nature of these highly occupied modes (outside of the
condensate itself) is not widely appreciated, and is the central topic
we address in this paper. We refer to these modes as the constituting
the \emph{classical region}, for which the classical field approximation
is appropriate: We can replace the quantum mode operators $\{\hat{a}_{j}\}$, satisfying the commutation relations $[\hat{a}_{i},\hat{a}_{j}^{\dagger}]=\delta_{ij}$,
with the c-number amplitudes $\{ c_{j}\}$. Modes outside the classical
region are more sparsely
 occupied and are poorly described by the classical
field approximation --- these modes constitute the \emph{incoherent
region.} In this paper we characterize the classical region of a trapped
Bose gas as a function of temperature and total particle number, and
show that the classical field approximation should be widely applicable
to current experiments.

 \begin{figure}
\includegraphics[%
  width=6.0cm,
  keepaspectratio]{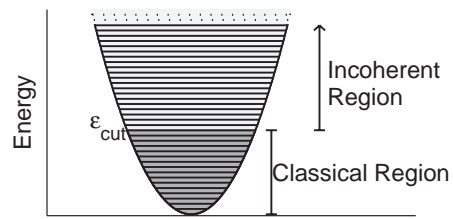}

\caption{\label{Fig:Classicalregion} Schematic diagram showing the classical
and incoherent regions of the single particle spectrum for a harmonically
trapped Bose gas. At the energy $\epsilon_{{\rm cut}}$ the average
occupation is  $n_{{\rm cut}}\sim3$ (see the text).}
\end{figure}

Schematically we show the classical and incoherent regions in
Fig.~\ref{Fig:Classicalregion}, for the case of the harmonically
trapped gas. 
The classical region is quantitatively defined as those single
particle modes with mean occupation greater than $n_{{\rm {cut}}}$,
where $n_{{\rm {cut}}}$ is the minimum occupation we require for
a mode to be designated as \emph{classical}. As the single particle
occupation monotonically decreases with increasing energy, the boundary of the
classical region occurs at an energy we define to be the \emph{cutoff energy,} $\epsilon_{{\rm cut}}$. With reference to the electromagnetic
case shown in Figs. \ref{Fig:Blackbody}(a)-(b), we see that
an adequate condition for the classical field description to be valid
is found by taking $n_{{\rm cut}}\sim3$ (i.e.\ blackbody modes with
a mean occupation of $\ge3$ photons are equally well-described by
the Planck and Rayleigh-Jeans distributions). We note that other authors
have suggested that taking $n_{{\rm cut}}\sim5$ to $10$ particles
may be more suitable \cite{Gardiner2003a}. The precise value of
$n_{{\rm cut}}$ used will be unimportant for the qualitative characteristics
of the classical region we investigate here.

\subsection{Projected Gross-Pitaevskii Equation Formalism}

In this paper we are not primarily concerned with the details of applying
the classical field technique to model Bose gases, but will briefly
review the formalism to establish the context and importance of the
classical region. The evolution equation for the classical modes of an
interacting Bose field
 is given by Projected Gross-Pitaevskii equation (PGPE) \cite{Davis2001a,Davis2001b,Davis2002a,Blakie2004}:
\begin{eqnarray}
i\hbar\frac{\partial\Psi}{\partial t} & = &
 \left(-\frac{\hbar^{2}}{2m}\nabla^{2}+V_{{\rm trap}}(\mathbf{x})\right)\Psi
 +\, \mathcal{P}\bigg\{ U_{0}|\Psi|^{2}\Psi\bigg\},\label{eq:PGPE}\end{eqnarray}
where $\Psi=\Psi(\mathbf{x},t)$ is the classical matter wave field
(i.e.\ describes the atoms in the classical region), $V_{{\rm trap}}(\mathbf{x})$
is the external trapping potential and $U_{0}=4\pi a\hbar^{2}/m$,
with $a$ the s-wave scattering length. The projector is defined as
\begin{equation}
\mathcal{P}\{ F(\mathbf{x})\}\equiv\sum_{n\in\mathcal{C}}\varphi_{n}(\mathbf{x)}\int d^{3}\mathbf{x}'\varphi_{n}^{*}(\mathbf{x'})F(\mathbf{x'}),\label{eq:projector}\end{equation}
where $\varphi_{n}(\mathbf{x})$ are eigenstates of the single particle
Hamiltonian $H_{{\rm sp}}=-\frac{\hbar^{2}}{2m}\nabla^{2}+
V_{{\rm trap}}(\mathbf{x})$
and the summation is restricted to modes in the classical region.
The action of $\mathcal{P}$ in Eq.~(\ref{eq:projector}) is to project
the arbitrary function $F(\mathbf{x})$ into the classical region.
A description of the incoherent particles will also be required for
quantitative comparison with experiments and a formalism for including
this into the PGPE theory has been presented in Ref. \cite{Gardiner2003a}.

Previous work has shown that the PGPE will evolve randomised initial conditions
to thermal equilibrium \cite{Davis2001a,Davis2002a,Blakie2004,
Goral2001a,Goral2002a}.  The equilibrium state that is reached depends on three input
parameters: (i) The number of atoms in the classical region, which
we refer to as $N_{{\rm below}}$; (ii) The total energy content of
the classical region $E$; (iii) The cutoff energy $\epsilon_{{\rm cut}}$
determining the size of the classical region (i.e.\ the number of classical
modes $M_{\rm{below}}$ \footnote{$M_{\rm{below}}$ is more appropriate than $\epsilon_{\rm{cut}}$ for identifying the appropriate projector ${\mathcal{P}}$ to use. This is because $\epsilon_{\rm{cut}}$ cab be shifted by interactions, whereas $M_{\rm{below}}$ is much less sensitive to this.}). 

In dynamical simulations of Eq.~(\ref{eq:PGPE}), interactions cause
the system to rapidly thermalize \cite{Davis2002a}, and subsequently
(using the ergodic hypothesis) time-averaging can be used to extract
equilibrium quantities. A major challenge for classical field approaches
has been to determine thermodynamics quantities that are derivatives
of entropy, such as temperature. This was recently overcome in 
Ref.~\cite{DavisTemp,Davis2005a}, using formalism initially 
developed by Rugh \cite{Rugh1997a}.
A remaining issue with using the classical field approach is that
the thermodynamic parameters for the simulation (such as the temperature,
and the total number of atoms in the system, i.e.\ including those
above the cutoff) are only apparent \emph{a posteriori}. The characterization
of the classical region we give here should facilitate making a reasonable
\emph{a priori} estimate of parameters for PGPE simulations and should
be of great practical benefit.

\section{Formalism}

For the results presented in Figs. \ref{Fig:CFregions} and \ref{Fig:CFcutoff}  we consider (bosonic) rubidium-87 in an isotropic harmonic trapping
potential with single-particle energy spectrum $\epsilon=\hbar\omega(n_{x}+n_{y}+n_{z}+\frac{3}{2}),$
where $\omega=2\pi\times100$ Hz is the harmonic trap frequency (chosen to be comparable with typical experiments). To account for interaction effects we use the semiclassical Hartree-Fock (HF) density of states
\begin{equation}
\rho_{\rm{HF}}(\epsilon)=\int\frac{d^3\mathbf{x}d^3\mathbf{p}}{(2\pi\hbar)^3}\delta\left(\epsilon-\left[\frac{p^2}{2m}+V_{\rm{trap}}(\mathbf{x})+2U_0n_c(\mathbf{x})\right]\right),\label{rhoHF}
\end{equation}
where the condensate density is given by the Thomas-Fermi approximation
\begin{equation}
n_c(\mathbf{x})=\left\{
   \begin{matrix} 
     \frac{1}{U_0}(\mu_{\rm{TF}}-V_{\rm{trap}}(\mathbf{x})),\qquad & \mu-V_{\rm{trap}}(\mathbf{x})\ge0, \\
      0, & \mu_{\rm{TF}}-V_{\rm{trap}}(\mathbf{x})<0, \\
   \end{matrix}
\right\}
\end{equation}
and $\mu_{\rm{TF}}$ is the Thomas-Fermi approximation to the condensate eigenvalue (e.g. see \cite{Dalfovo1999,Bijlsma2000a}). The Thomas-Fermi approximation is only valid for large condensate number, however, for small condensate numbers the density of states is largely unchanged from the ideal gas case, and so for our purposes  this approximation is valid for all temperatures.

To study the statistical properties of this system at finite temperature
we work in the grand canonical ensemble thermally occupying the  particle states described by (\ref{rhoHF}) using the Bose-Einstein distribution function\begin{equation}
n_{{\rm BE}}(\epsilon)=\frac{1}{e^{\beta\epsilon}/z-1},\label{eq:BEdist}\end{equation}
 where $\beta=1/k_BT$ is the inverse temperature,  $z = e^{\beta \mu}$ is the
fugacity, and the excitation energy $\epsilon$ is measured relative to the chemical potential ($\mu_{\rm{TF}}$).  To calculate the thermodynamic properties as a function of the total
number of particles $N_{T}$  it is necessary
to find the fugacity subject to the requirement that the
mean number of particles is equal to $N_{T}$. This parameter must be adjusted carefully as  the condensate number, condensate density and excitation spectrum all change with $z$.

Of interest for us is to identify the properties of the classical region, in particular the number of classical  modes ($M_{\rm{below}}$) and their combined occupation  ($N_{\rm{below}}$). 
According to Eq. (\ref{eq:BEdist}) the energy at which the mean occupation is equal to $n_{\rm{cut}}$ is given by 
\begin{equation}
\epsilon_{{\rm cut}}=k_BT\ln\left[z\left(1+\frac{1}{n_{{\rm cut}}}\right)\right], \label{eq:ecut}\end{equation}
Integrating the density of states (\ref{rhoHF}) up to this energy we obtain the number of particle states,  
\begin{equation}
M_{\rm{below}}=\int_0^{\epsilon_{\rm{cut}}}d\epsilon\,\rho_{\rm{HF}}(\epsilon),
\end{equation} 
which we refer to as the number of classical modes. Finally, the combined occupation of the classical modes is given by
\begin{equation}
N_{\rm{below}}=\int_0^{\epsilon_{\rm{cut}}}d\epsilon\,n_{\rm{BE}}(\epsilon)\rho_{\rm{HF}}(\epsilon).
\end{equation} 
\section{Results}
\subsection{Ideal Gas}
\begin{figure}
\includegraphics[%
  width=8.5cm,
  keepaspectratio]{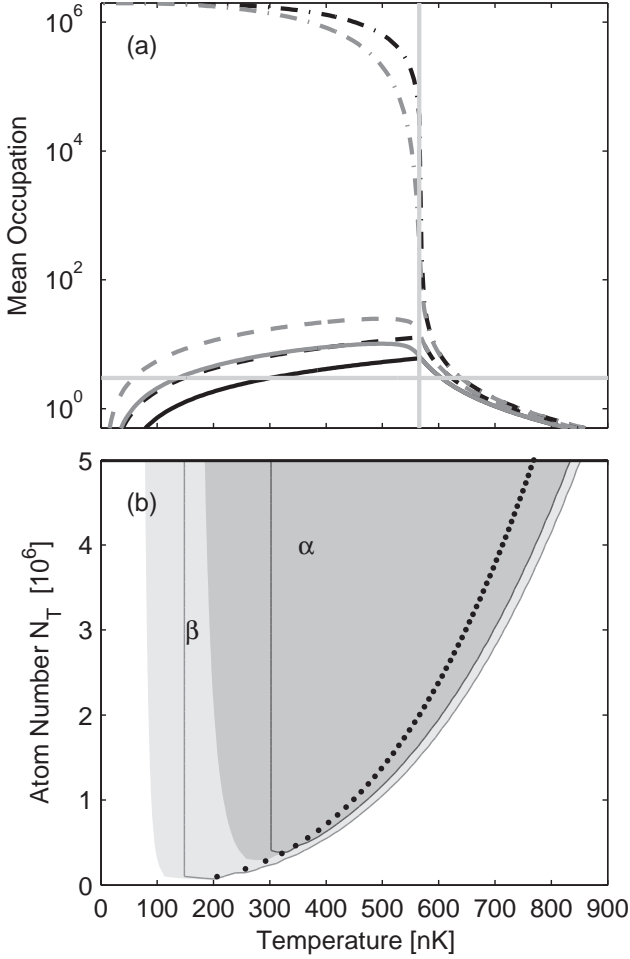}

\caption{\label{Fig:CFregions} (a) Mean occupation of the condensate and selected excited single particle states for both ideal (black curves) and interacting (grey curves) systems of $2\times10^6$ rubidium-87 atoms in a 3D 100 Hz isotropic trap. Condensate: dash-dot. $M=120^{\rm{th}}$ single particle level: solid. $M=1000^{\rm{th}}$ single particle level: dashed. 
The horizontal  light-grey line
indicates $n_{{\rm cut}}=3$ particles per mode. The vertical light-grey line marks $T_c$ for 
this system. (b) Phase diagram of the classical region. The
dark shaded region labelled $\alpha$ is where the
$M=1000{}^{{\rm th}}$ single particle level 
has a mean occupation of greater than 3 particles. The
lighter shaded region labelled $\beta$ is where
the $M=120{}^{{\rm th}}$ single particle level 
has a mean occupation of greater than 3 particles. The boundaries of these regions for the non-interacting case are shown as grey lines and $T_c$ is shown as with black dots. }
\end{figure}

Typical results for the mode occupations are shown in Fig.~\ref{Fig:CFregions}(a)
for a system of $N_{T}=2\times10^{6}$ atoms. Here we discuss the ideal gas results, and refrain from considering the interacting results until the next subsection.

 Far above the transition
temperature, $T_{c}\approx566$ nK, the occupation of all modes is
small compared to unity and there is no classical region. At $T\approx640$ nK
the occupation of several modes exceed $n_{{\rm cut}}=3$ atoms and
a classical region develops in the system. At temperatures equal to,
or less than $T_{c}$, the occupation of the condensate mode dominates
the system, however a large number of other modes have occupations
exceeding $n_{{\rm cut}}$. At temperatures near $T_{c}$
of order a thousand single particle states participate in the
classical region. This can be seen from Fig.~\ref{Fig:CFregions}(a)
where the where the dashed line
representing the occupation of the $1000^{{\rm th}}$ single particle
state is above the dotted line marking $n_{{\rm cut}}$ for a range of temperatures encompassing $T_c$.
States of lower energy remain highly occupied over a broader temperature
range (e.g. the $\epsilon=8.5\hbar\omega$ mode occupation, represented
by the dash-dot line in Fig.~\ref{Fig:CFregions}(a), exceeds $n_{{\rm cut}}$
over a broader range than the $\epsilon=17.5\hbar\omega$ mode). Thus the
number of modes in the classical region $M_{\rm{below}}$   varies significantly with temperature, but is largest at $T_c$  where the
non-condensate modes have their maximum occupation.  

The classical regions for two particular modes are shown in Fig.~\ref{Fig:CFregions}(b)
as a function of temperature and total number of particles. In region
$\alpha$  the mean occupation of the $\epsilon=17.5\hbar\omega$
state exceeds $n_{{\rm cut}}$. In the broader region $\beta$  the mean occupation of the $\epsilon=8.5\hbar\omega$ state
exceeds $n_{{\rm cut}}$. The left hand edges of the regions are independent
of  $N_{T}$ (i.e.\ the boundary of the region is vertical). This is
because the thermal cloud is saturated for $T<T_{c}$, and the mode
occupation only depends on temperature. We can quantitatively predict
the left-hand boundary by noting that below transition temperature
the chemical potential is well approximated by taking it equal to
the ground state energy, i.e.\ $\mu \approx \epsilon_{0}$, inverting the Bose-Einstein distribution function we obtain  \begin{equation}
T_{LH}=\frac{\epsilon-\epsilon_{0}}{k_B\ln(1+1/n_{{\rm cut}})},\label{eq:Tcut}\end{equation} independent of $N_T$ and in agreement with the left-hand boundaries in Fig.~\ref{Fig:CFregions}(b). 

The right hand boundary occurs above $T_{c}$ where $\mu$ is different
from $\epsilon_{0}$, and is seen to be dependent on $N_{T}$. We
note that for large $N_{T}$ the classical region extends to temperatures
considerably above $T_{c}$, indeed for the case considered in Fig.~\ref{Fig:CFregions}(b) the classical region begins $\sim100$ nK above
$T_{c}$ for $N_{T}>3\times10^{6}$ particles.

\begin{figure}
\includegraphics[%
  width=8.5cm,
  keepaspectratio]{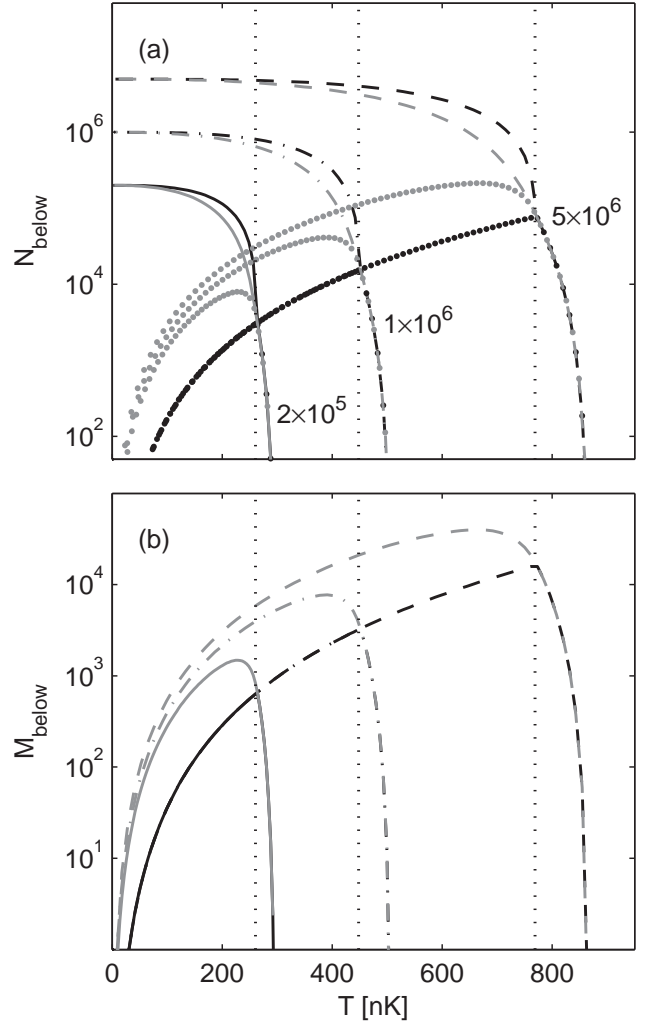}

\caption{\label{Fig:CFcutoff} The classical region for rubidium-87 atoms in a 3D isotropic harmonic trap with $\omega=2\pi\times100$ Hz. (a) The total number
of atoms in modes with occupation greater than 3 ($N_{{\rm below}}$)
for systems with a total of $5\times10^{5}$ (solid line), $1\times10^{6}$
(dash-dot line), and $5\times10^{6}$ (dashed line) atoms. The dotted curves show the number of atoms constituting $N_{{\rm below}}$ that
are not in the condensate. (b) The number of modes in the classical region
$M_{{\rm below}}$  for the systems in (a). 
For  all curves in this plot black curves are for the non-interacting case and  grey curves include the effects of interactions. For reference $T_{c}$ is indicated
using vertical dotted lines in each plot.}
\end{figure}

In Fig.~\ref{Fig:CFcutoff} we characterise the total number of atoms
 ($N_{{\rm below}}$) and modes ($M_{{\rm below}}$) participating in the classical region as a function of temperature
for several values of $N_{T}$. For $T<T_{c}$, $N_{{\rm below}}$
is dominated by the condensate mode and scales as $N_{{\rm {\rm below}}}\sim\{1-(T/T_{c})^{3}\}$,
as can be seen for the three cases of $N_{T}$ considered in Fig.~\ref{Fig:CFcutoff}(a). The number of classical region particles not
in the condensate is shown as the dotted curve in Fig.~\ref{Fig:CFcutoff}(a). We notice that up to the transition temperature this curve is
independent of $N_{T}$, due to the saturated nature of the thermal
cloud. Though the population of these classical non-condensate particles
is much less than the condensate, for the cases considered in 
Fig.~\ref{Fig:CFcutoff}(a) we see that they may constitute 
up to $~\sim10^5$ particles. Additionally, 
above $T_{c}$ there is no condensate and
the classical non-condensate particles dominate  the classical region.

In Fig.~\ref{Fig:CFcutoff}(b) we consider the behaviour of the cutoff energy. For $T<T_{c}$ the cutoff energy increases linearly with
temperature, and is independent of $N_{T}$. As discussed above, for
$T<T_{c}$ the chemical potential is well-approximated as the ground state
energy $\mu\to\epsilon_{0}$, and we can find an explicit expression
for the cutoff energy in terms of the temperature\begin{equation}
\epsilon_{{\rm cut}}=k_BT\ln\left(1+\frac{1}{n_{{\rm cut}}}\right)+\epsilon_{0},\qquad(T\le T_{c}),\label{eq:ecut2}\end{equation}
(or equivalently, $M_{\rm{below}}=\epsilon_{\rm{cut}}^3/6(\hbar\bar\omega)^3$, where $\bar\omega$ is the geometric mean of the harmonic trap frequencies). 
 Reaching its maximum value at $T_{c}$, the cutoff energy decreases
for $T>T_{c}$ as the classical region shrinks due to the rapid decrease
in $\mu$.

\subsection{Hartree-Fock Results}
The results for the interacting gas, calculated using the density of states given in Eq. (\ref{rhoHF}), are also shown in Figs.~\ref{Fig:CFregions} and~\ref{Fig:CFcutoff}. 
There are many qualitative differences introduced in the interacting case.
  First, for the interacting system the size and number of particles in the classical regions tends to be larger than the ideal case. Second, the number of modes of the classical region ($M_{\rm{below}}$) tends to peak at temperatures below $T_c$ for the interacting case, whereas it peaks at $T_c$ for the ideal case. These effects are related and can be understood by examining the density of states (\ref{rhoHF}). Due to the mean field repulsion of the condensate, the lowest energy excitations do not occupy the phase space coordinates near the origin,  but for position values near the surface of the condensate. Because of the increased amount of phase space available at the surface of the condensate more low energy states are available. As the temperature increases the condensate fraction and radius both decrease, reducing this phase space enhancement. The observed maximum of  $M_{\rm{below}}$ in Fig.~\ref{Fig:CFcutoff}(b) occurs due to the competition between increasing temperature (which tends to increase the number of accessible states) and reducing condensate radius (which reduces the number of low energy modes).

\section{Comparison to PGPE}
It is of interest to assess how well the ideal gas and Hartree-Fock results for the classical region describe the classical field case. Because the PGPE approach includes interactions between the excitations  we expect quantitative differences with the Hartree-Fock results to arise. Our procedure for calculating equilibrium properties for the trapped Bose gas is discussed in Ref. \cite{Davis2006a}. Briefly, the PGPE system is a microcanonical system with external constants of motion (i.e. thermodynamic constraints) of $N_{below}$ and total energy $E$. Additionally, to avoid an ultraviolet divergence a restriction in the single particle modes must be made by selecting $M_{below}$. After equilibrium is found and the properties of the classical region are determined, such as mean density, temperature and chemical potential, the incoherent region can be found using a semiclassical Hartree-Fock analysis.

\begin{figure}
\includegraphics[%
  width=8.5cm,  keepaspectratio]{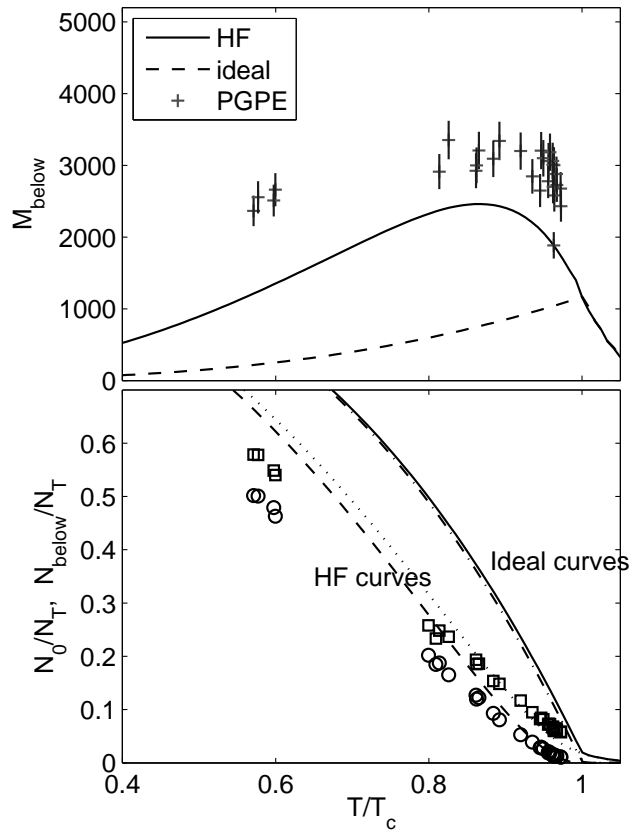}

\caption{\label{Fig:PGPE}  Comparison of classical region estimates using the methods described in this paper to the 
numerical PGPE results as a function of temperature. (a) Number of classical region modes for $n_{\rm{cut}}=2.8$: HF (solid), ideal gas (dashed), PGPE solution (+), indicating estimated error bars vertically. (b) Classical region fractional population: HF (dotted), ideal gas (solid), PGPE solution (square). Condensate fraction: HF (dashed), ideal gas (dash-dot), PGPE solution (circle).
Harmonic trap with oscillation frequencies $\{f_x,f_y,f_z\}=\{1,1,\sqrt{8}\}\times40$ Hz containing approximately $3\times10^5$ $^{87}$Rb atoms.
}
\end{figure}

The results of this study are shown in Fig. \ref{Fig:PGPE}.  These confirm that the HF predictions are qualitatively in much better agreement with the PGPE theory than the ideal case. We note the following points:
\begin{enumerate}
\item Temperature in Fig.~\ref{Fig:PGPE} has been scaled in terms of the ideal transition temperature. The PGPE exhibits the condensation transition at $T/T_c<1$ due to the downward shifts in critical temperature arising from finite size and interaction effects (neither of these effects are in the ideal or HF results).
\item The PGPE results suggest that number of classical modes, $M_{\rm{below}}$, does not increase up to the critical point (as predicted by the ideal result), and are consistent with the HF prediction of a maximum value occurring for $T<T_c$.
\item The number of classical region modes is considerably larger than predicted by the HF method.  However, as the total number of modes $M_{\rm{below}}$ is  proportional to $\epsilon_{cut}^3$  for the ideal gas then the increase in $\epsilon_{cut}$ is not as significant.

\end{enumerate}

\section{Conclusions and Outlook}

We have characterized the classical region, containing modes with
occupations higher than $n_{{\rm cut}}\sim3$, as a function of the
total number of particles and temperature for a harmonically trapped
Bose gas. To do this we have examined the number of modes and number
of particles in the classical region, and the behaviour of the energy
cutoff. 

For the ideal case we have shown that for $T>T_{c}$ the cutoff energy is dependent
on the total number of atoms in the system and the temperature, whereas
for $T<T_{c}$ the cutoff energy increases linearly with temperature
and is independent of $N_{T}$.

We have studied interaction effects using a Hartree-Fock approach which we have compared with full PGPE simulations. Our results show that interactions have a considerable effect on the classical region, and appear to make the size of the classical region much greater than predicted by an ideal gas analysis. 

In general  we have demonstrated that for systems with of order a million
particles the classical region may consist of thousands of single particles states  and extend to temperatures of order $100$ nK above $T_{c}$. Overall these results demonstrate that the PGPE technique has a wide range of applicability to trapped Bose gases and should facilitate better parameter estimation necessary for effectively applying the PGPE formalism to model experiments.

\acknowledgments  PBB would like to acknowledge the use of classical field data provided by A. Bezett   and thanks the Marsden Fund of New Zealand  for supporting this research.

\bibliographystyle{apsrev}
\bibliography{projector} 

\end{document}